# Utilizing Machine Learning to Greatly Expand the Range and Accuracy of Bottom-Up Coarse-Grained Models Through Virtual Particles

Patrick G. Sahrmann, Timothy D. Loose, Aleksander E.P. Durumeric, and Gregory A. Voth*

*Department of Chemistry, Chicago Center for Theoretical Chemistry, James Franck Institute, and Institute for Biophysical Dynamics, The University of Chicago, Chicago, IL 60637, USA*

*Corresponding Author: gavoth@uchicago,edu

ABSTRACT

Coarse-grained (CG) models parameterized using atomistic reference data, i.e., "bottom up" CG models, have proven useful in the study of biomolecules and other soft matter. However, the construction of highly accurate, low resolution CG models of biomolecules remains challenging. We demonstrate in this work how virtual particles, CG sites with no atomistic correspondence, can be incorporated into CG models within the context of relative entropy minimization (REM) as latent variables. The methodology presented, variational derivative relative entropy minimization (VD-REM), enables optimization of virtual particle interactions through a gradient descent algorithm aided by machine learning. We apply this methodology to the challenging case of a solvent-free CG model of a 1,2-dioleoyl-sn-glycero-3-phosphocholine (DOPC) lipid bilayer and demonstrate that introduction of virtual particles captures solvent-mediated behavior and higher-order correlations which REM alone cannot capture in a more standard CG model based only on the mapping of collections of atoms to the CG sites.

**I. Introduction**

Atomistic molecular dynamics (MD) simulations have enabled key insights into biological and material processes.[1-3] but modern hardware limits the practicality of MD to the study of millions of atoms on the multi-microsecond scale. This precludes sampling of many biologically relevant phenomena, such as macromolecular assembly.[4] Coarse-grained (CG) models aim to extend the spatiotemporal scales of MD by simulating a lower-resolution representation of a system, increasing the efficiency of the simulation.[5-8]

CG model construction and parameterization often follow either a top-down or bottom-up approach.[7, 8] In the top-down approach, a CG model is parameterized to directly reproduce macroscopic properties such as thermodynamic data. Popular top-down approaches such as MARTINI have been used to simulate biomolecular structures such as proteins and multicomponent lipid bilayers.[9-11] However, these models are not parameterized to reproduce the microscopic correlations and enthalpy-entropy decompositions underpinning these properties.[12] Bottom-up models instead aim to reproduce the microscopic behavior of a reference atomistic model, with the intent of indirectly capturing emergent behavior.[8, 13-18] While bottom-up CG models have an explicit correspondence to the atomistic representation and in principle have a greater potential in accurately describing the underlying physics, their application is limited due to the necessity of imperfect basis sets (force field expressions) to represent the CG interactions, as well as issues of representability and transferability.[19-21]

Ideally, equilibrium simulations of bottom-up CG models will reproduce the configurational distribution of the CG variables implied by the atomistic reference simulation. The exact CG model Hamiltonian whose Boltzmann statistics reproduce the reference distribution is referred to as the CG variable potential of mean force (mbPMF),[14, 15] and is often many-body in nature.[22,23] In the limit of infinite sampling and a perfect basis set to represent the CG interactions, bottom-up CG methods such as Multiscale Coarse-graining[13, 15, 16] (MS-CG) and Relative Entropy Minimization[17, 18, 24] (REM) are guaranteed to reproduce the mbPMF. However, practical considerations often relegate the CG force-field (basis set) to pairwise non-bonded interactions. Enhancing the expressivity of CG force fields beyond a pairwise basis set through explicit higher-order terms[25] and order parameter (e.g., local density) based interactions[26-30] enables the capture of certain many-body statistics. However, for biological systems, it is generally not clear which higher-order terms should be included. Machine-learned CG force-fields can be utilized to construct general approximations to many-body statistics,[31-35] albeit at increased computational cost.

Practical implementations of bottom-up CG methods possess characteristic approximations often affiliated with matching lower-order correlations. Certain methods such as REM,[24] inverse Monte Carlo[36] (IMC), and iterative Boltzmann inversion[37] (IBI) guarantee fidelity of correlations which correspond to a particular interaction in the CG force field (e.g., optimization of a pair interaction reproduces the corresponding pair correlation). This matching of lower-order correlations often comes at the expense of accurately describing the higher-order correlations. The MS-CG method instead attempts a complete fit to the mbPMF through approximation of the mean forces on CG sites via an integral equation connection of the two- and three-body correlations,[14] but it is not guaranteed to completely reproduce any structural correlations when an incomplete basis set is used to represent the interactions at the CG level.[38] For any CG method, improvement of the CG models must proceed through either an increase in the complexity of the CG force field or a change in the considered CG resolution.

Recent advances have, however, focused on augmenting the CG force field while maintaining a pairwise basis set. For example, the pairwise interactions in CG models can be altered to explicitly include higher-order correlations by projecting many-body interactions onto the pairwise basis set.[39] Alternatively, various kinds of "virtual" sites have been introduced in bottom-up CG models, which can reintroduce orientational information in isotropic CG models and/or capture solvent-mediated interactions.[40-42]

One can define virtual sites generally as sites with more complex relationships to the CG (or AA) model than other sites within the model. Such virtual sites may be related to the atomistic system through a non-linear mapping, or they may be defined in relation to other variables present in the model system. Ideally, the optimization of virtual site interactions in a CG model results in direct improvement of the behavior of the non-virtual or "real" CG particles. Consequently, the inclusion of virtual particles into a CG model may be most fruitful for very low-resolution CG models, as shown in Figure 1. Virtual sites have similarly been introduced in both all-atom (AA) and top-down CG models to suit a variety of model-specific needs. These include imparting anisotropic projections, such as in the TIP4P water model, improving stability of cholesterol in top-down CG models, and aiding in mixed resolution AA/CG simulations.[43-45]

In the context of bottom-up CG models, development of virtual particles has generally been impeded by their complex relationship to the associated CG model. Methods centered around distribution matching, e.g., REM, can be directly incorporated so long as virtual sites are observed in the reference simulation, and can furthermore indirectly enable force-based parameterization.[42] However, parameterization of virtual sites directly through force-based methods, although possible for particularly defined virtual sites,[41] is limited. In either case, these limitations are a consequence of missing information regarding virtual site behavior within the reference (usually all-atom) simulation. Such missing information may include forces for virtual sites whose non-linear mapping precludes their direct computation or, for latent virtual sites, a reference distribution altogether. We consider virtual sites of the latter form in this manuscript.

Although limited, certain methods do exist which are capable of optimizing CG models for which a reference virtual site distribution is not directly available. Such methods include Inverse Monte Carlo (IMC) and Adversarial-Residual-Coarse-Graining (ARCG).[46] In IMC, the expectation of a vector-valued observable, necessarily a function of exclusively real particles, is considered for optimization.[36] ARCG by contrast minimizes a selected *f*-divergence between reference and model distributions at the CG resolution, which subsequently results in the optimization of a scalar-value observable whose form must be obtained variationally.[46]

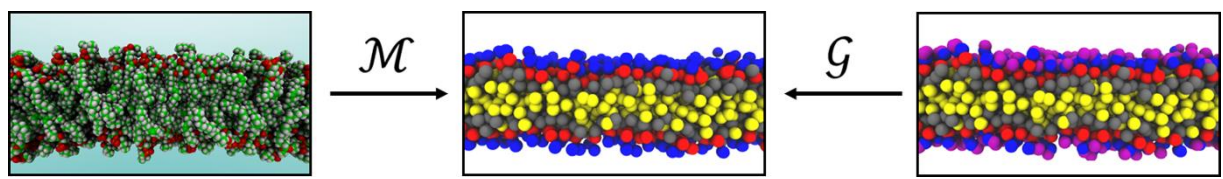

**Figure 1.** Example of CG mapping and VCG mapping for a lipid bilayer. The mapping operator $\mathcal{M}$ relates the solvated AA bilayer (left) to the implicit solvent CG resolution (middle), while the mapping operator $\mathcal{G}$ relates the VCG resolution (right) which contains virtual particles (purple) to the CG resolution.

Although the two methods are conceptually different, their approach to virtual particle optimization is similar in that the observable expectations considered must be calculable either directly (IMC) or indirectly through variational search (ARCG) at the CG resolution.

We present here a REM-based methodology for virtual particle optimization, variational derivative relative entropy minimization (VD-REM), which utilizes a distinct variational statement to approximate the derivative of the relative entropy. Similar to both IMC and ARCG, this enables a gradient descent algorithm for which all calculations are performed at the CG resolution. We find that machine-learned models can be useful in VD-REM in approximating target functions necessary for optimization. We apply VD-REM to construct a VCG model of a 1,2-dioleoyl-sn-glycero-3-phosphocholine (DOPC) lipid bilayer without solvent (i.e., "solvent free" or sometimes called "implicit" solvent). Inclusion of virtual particles optimized by VD-REM demonstrates improved model behavior at the CG resolution, including recapture of self-assembly and bilayer flexibility. Lastly, we note that the presented methodology bears conceptual similarities with Predictive Coarse-Graining,[47] which further suggests an alternative approach to virtual particle optimization, as detailed in the Discussion.

## II. Theory

### A. Model Definitions

Before introducing virtual CG particles, we first explicitly define the AA and CG models. The fine-grained (FG) model is described by a collection of masses, **m**, and configurational and momentum degrees of freedom $\mathbf{r}^n$ and $\mathbf{p}^n$, respectively. Reference simulations of the FG model are generated using the dynamics of the FG Hamiltonian

$$h(\mathbf{r}^n, \mathbf{p}^n) = u(\mathbf{r}^n) + \sum_{i=1}^{n} \frac{\mathbf{p}_i^2}{2m_i} \tag{1}$$

In the canonical ensemble, molecular dynamics simulations sample from the Boltzmann distribution generated by the FG Hamiltonian

$$p^{\mathrm{FG}}(\mathbf{r}^n, \mathbf{p}^n) = \frac{1}{z} \exp(-\beta h(\mathbf{r}^n, \mathbf{p}^n)) \tag{2}$$

where $\beta = \frac{1}{k_B T}$. The normalization factor (ignoring constants due to indistinguishability and unit normalization) is the classical partition function, and can be written as

$$z = \iint e^{-\beta h(\mathbf{r}^n, \mathbf{p}^n)} d\mathbf{r}^n d\mathbf{p}^n = z_\mathbf{r} z_\mathbf{p} \tag{3}$$

The separability of position and momentum implies the configurational and momentum marginal densities can be expressed as

$$p_\mathbf{r}^{\mathrm{FG}}(\mathbf{r}^n) = \frac{1}{z_\mathbf{r}} \exp(-\beta u(\mathbf{r}^n)) \tag{4}$$

$$p_\mathbf{p}^{\mathrm{FG}}(\mathbf{p}^n) = \frac{1}{z_\mathbf{p}} \exp\left\{-\beta \sum_{i=1}^{n} \frac{\mathbf{p}_i^2}{2m_i}\right\} \tag{5}$$

As in prior development in MS-CG methodology, we define a mapping operator, $\mathcal{M} = (\mathcal{M}_r, \mathcal{M}_p)$, which connects the CG degrees of freedom, denoted $\mathbf{R}^N$ and $\mathbf{P}^N$, and FG degrees of freedom. We consider the mapping operator linear for both position and momentum coordinates such that

$$\mathbf{R}_I = \mathcal{M}_{\mathbf{r},I}(\mathbf{r}^n) = \sum_{i=1}^{n} c_{Ii} \mathbf{r}_i \tag{6}$$

$$\mathbf{P}_I = \mathcal{M}_{\mathbf{p},I}(\mathbf{p}^n) = \sum_{i=1}^{n} c_{Ii} \mathbf{p}_i \tag{7}$$

Additionally, we restrict the form of the mapping operator such that each FG particle is uniquely associated with a single CG site. The mapping between the FG and CG variables enables the construction of a probability density over the CG variables, i.e.

$$p_{\mathcal{M}_\mathbf{r}}^{\mathrm{ref}}(\mathbf{R}^N) = \int p_\mathbf{r}^{\mathrm{FG}}(\mathbf{r}^n) \delta(\mathcal{M}_\mathbf{r}(\mathbf{r}^n) - \mathbf{R}^N) \mathrm{d}\mathbf{r}^n \tag{8}$$

$$p_{\mathcal{M}_\mathbf{p}}^{\mathrm{ref}}(\mathbf{P}^N) = \int p_\mathbf{p}^{\mathrm{FG}}(\mathbf{p}^n) \delta(\mathcal{M}_\mathbf{p}(\mathbf{p}^n) - \mathbf{P}^N) \mathrm{d}\mathbf{p}^n \tag{9}$$

A CG model which samples from these distributions will reproduce the equilibrium static behavior of the FG model, at the CG representation, with complete fidelity. To this end, a model Hamiltonian, $H_\mathcal{M}$, with parameters $\boldsymbol{\theta}$, is introduced

$$H_\mathcal{M}(\mathbf{R}^N, \mathbf{P}^N; \boldsymbol{\theta}) = U_\mathcal{M}(\mathbf{R}^N; \boldsymbol{\theta}) + \sum_{I=1}^{N} \frac{\mathbf{P}_I^2}{2M_{I,\mathcal{M}}} \tag{10}$$

The model Hamiltonian generates its own equilibrium density, i.e.

$$p_\mathcal{M}^{\mathrm{mod}}(\mathbf{R}^N, \mathbf{P}^N; \boldsymbol{\theta}) = \frac{1}{Z} \exp(-\beta H_\mathcal{M}(\mathbf{R}^N, \mathbf{P}^N; \boldsymbol{\theta})) \tag{11}$$

The model equilibrium density can again be separated into position and momentum distributions, $p_\mathcal{M}^{\mathrm{mod}}(\mathbf{R}^N, \mathbf{P}^N) = p_{\mathcal{M}_\mathbf{R}}^{\mathrm{mod}}(\mathbf{R}^N) p_{\mathcal{M}_\mathbf{P}}^{\mathrm{mod}}(\mathbf{P}^N)$. This defines the following consistency conditions:

$$p_{\mathcal{M}_\mathbf{R}}^{\text{mod}}(\mathbf{R}^N; \boldsymbol{\theta}) = p_{\mathcal{M}_\mathbf{r}}^{\text{ref}}(\mathbf{R}^N) \tag{12}$$

$$p_{\mathcal{M}_\mathbf{P}}^{\text{mod}}(\mathbf{P}^N) = p_{\mathcal{M}_\mathbf{p}}^{\text{ref}}(\mathbf{P}^N) \tag{13}$$

Momentum space consistency is trivially satisfied by a choice of mass related to the mapping operator.[15] Complete configurational consistency would necessitate introduction of many-body terms into the CG FF. Instead, bottom-up CG methods aim to obtain a parameter set $\boldsymbol{\theta}^\dagger$ which satisfies a minimization principle unique to each method.

### B. Variational Derivative Relative Entropy Minimization

At this point, we now consider the introduction of virtual particles into the CG model to produce a VCG model. The VCG model is described by configurational and momentum degrees of freedom $\hat{\mathbf{R}}^\nu$ and $\hat{\mathbf{P}}^\nu$, respectively. The parameters of the VCG model are separated into those describing interactions between exclusively real particles, $\boldsymbol{\theta}_c$, and those which feature at least one virtual particle, $\boldsymbol{\theta}_\nu$. We define the VCG Hamiltonian, $H_\mathcal{G}$, as

$$H_\mathcal{G}(\hat{\mathbf{R}}^\nu, \hat{\mathbf{P}}^\nu; \boldsymbol{\theta}_c, \boldsymbol{\theta}_\nu) = U_\mathcal{G}(\hat{\mathbf{R}}^\nu; \boldsymbol{\theta}_c, \boldsymbol{\theta}_\nu) + \sum_{I=1}^{\nu} \frac{\hat{\mathbf{P}}_I^2}{2M_{I,\mathcal{G}}} \tag{14}$$

The Boltzmann distribution generated by the VCG Hamiltonian is referred to as $p_{\text{pre}}^{\text{mod}}$. We consider a mapping operator, $\mathcal{G}$, which takes the VCG resolution to the CG resolution. The marginal densities of the VCG resolution are then connected to the associated CG densities through this mapping operator

$$p_{\mathcal{M}_\mathbf{R}}^{\text{mod}}(\mathbf{R}^N; \boldsymbol{\theta}_c, \boldsymbol{\theta}_\nu) = \int p_{\text{pre}}^{\text{mod}}(\hat{\mathbf{R}}^\nu; \boldsymbol{\theta}_c, \boldsymbol{\theta}_\nu) \delta(\mathcal{G}_{\hat{\mathbf{R}}}(\hat{\mathbf{R}}^\nu) - \mathbf{R}^N) d\hat{\mathbf{R}}^\nu \tag{15}$$

$$p_{\mathcal{M}_\mathbf{P}}^{\text{mod}}(\mathbf{P}^N) = \int p_{\text{pre}}^{\text{mod}}(\hat{\mathbf{P}}^\nu) \delta(\mathcal{G}_{\hat{\mathbf{P}}}(\hat{\mathbf{P}}^\nu) - \mathbf{P}^N) d\hat{\mathbf{P}}^\nu \tag{16}$$

We note that consistency in momentum space is trivially achieved for all real particles by setting $M_{I,\mathcal{G}} = M_{I,\mathcal{M}} \ \forall \ I$.[46] We note that the dependence of the marginal density $p_{\mathcal{M}_\mathbf{R}}^{\text{mod}}$ on the VCG model parameters $\boldsymbol{\theta}_c$ and $\boldsymbol{\theta}_\nu$ as shown in Equation 15 is not directly considered in the work presented and omit representation of this dependence via notation for clarity in future equations. Since all virtual particle degrees of freedom are removed prior to model evaluation, momentum consistency is achieved for virtual particles at any chosen mass.

Similarly, we introduce a mapping operator $\mathscr{g}$, which takes the unobserved VAA ensemble, $p_{\text{pre}}^{\text{ref}}$, with FG degrees of freedom $\hat{\mathbf{r}}^\mu$ and $\hat{\mathbf{p}}^\mu$, to the observed AA ensemble

$$p_{\mathbf{r}}^{\text{FG}}(\mathbf{r}^n) = \int p_{\text{pre}}^{\text{ref}}(\hat{\mathbf{r}}^\mu)\delta(\mathscr{g}_{\hat{\mathbf{r}}}(\hat{\mathbf{r}}^\mu) - \mathbf{r}^n)d\hat{\mathbf{r}}^\mu \tag{17}$$

$$p_{\mathbf{p}}^{\text{FG}}(\mathbf{p}^n) = \int p_{\text{pre}}^{\text{ref}}(\hat{\mathbf{p}}^\mu)\delta(\mathscr{g}_{\hat{\mathbf{p}}}(\hat{\mathbf{p}}^\mu) - \mathbf{p}^n)d\hat{\mathbf{p}}^\mu \tag{18}$$

As the remainder of this article is concerned only with configurational consistency, we omit subscripts on mapping operators $\mathcal{M}$, $\mathcal{G}$, and $\mathscr{g}$ as the type of mapping is implicitly understood to be configurational. We denote the collection of virtual particles $\mathbf{Y}$ such that $\hat{\mathbf{R}}^\nu = (\mathbf{R}^N, \mathbf{Y})$ and $\hat{\mathbf{r}}^\mu = (\mathbf{r}^n, \mathbf{Y})$. The relative entropy between the configurational densities when including virtual particles is then

$$S_{\text{rel}} = \int p_{\text{pre}}^{\text{ref}}(\mathbf{r}^n, \mathbf{Y}) \ln\left(\frac{p_{\text{pre}}^{\text{ref}}(\mathbf{r}^n, \mathbf{Y})}{p_{\text{pre}}^{\text{mod}}(\mathcal{M}(\mathbf{r}^n), \mathbf{Y})}\right) d\mathbf{r}^n d\mathbf{Y} + \langle S_{\text{map}} \rangle_{\text{VAA}} \tag{19}$$

We note that the introduction of virtual particles does not affect the ensemble averaged value of the mapping entropy, which can be evaluated equivalently in the AA ensemble. We next consider the derivative of the relative entropy with respect to a VCG model parameter

$$\frac{\partial S_{\text{rel}}}{\partial \theta} = \beta \left\langle \frac{\partial U_\mathcal{G}}{\partial \theta} \right\rangle_{\text{VAA}} - \beta \left\langle \frac{\partial U_\mathcal{G}}{\partial \theta} \right\rangle_{\text{VCG}} \tag{20}$$

For model parameters which govern interactions between exclusively real particles, the relative entropy derivative can be calculated explicitly by integration of the virtual particle degrees of freedom in the ensemble averages of Equation 20. This produces the standard relative entropy derivative

$$\frac{\partial S_{\text{rel}}}{\partial \theta_c} = \beta \left\langle \frac{\partial U_\mathcal{G}}{\partial \theta_c} \right\rangle_{\text{AA}} - \beta \left\langle \frac{\partial U_\mathcal{G}}{\partial \theta_c} \right\rangle_{\text{CG}} \tag{21}$$

The ensemble averages in Equation 21 can simply be evaluated over the configuration of real particles, just as in a standard REM algorithm.[24] We note that functional minimization of the relative entropy proceeds identically for real particle functions as in standard REM, and hence the recapitulation of the correlations dual to the potential energy function optimized are also guaranteed for VD-REM.

The relative entropy derivative with respect to a model parameter that governs virtual particle interactions cannot be calculated in the same manner as the VAA ensemble, unlike the VCG ensemble, is not observed. We proceed by defining the following conditional distribution

$$p_{\text{pre}}^{\text{mod}}(\widehat{\mathbf{R}}^\nu) = p_{\mathcal{M}_{\mathbf{R}}}^{\text{mod}}(\mathbf{R}^N) p^{\text{cond}}(\mathbf{Y}|\mathbf{R}^N; \boldsymbol{\theta}_c, \boldsymbol{\theta}_v) \qquad (22)$$

$$p_{\text{pre}}^{\text{ref}}(\widehat{\mathbf{r}}^\mu) = p_{\mathbf{r}}^{\text{FG}}(\mathbf{r}^n) p^{\text{cond}}(\mathbf{Y}|\mathcal{M}(\mathbf{r}^n); \boldsymbol{\theta}_c, \boldsymbol{\theta}_v) \qquad (23)$$

We note that the conditional distributions in Equations 22 and 23 are identical, i.e., $p^{\text{cond}}(\mathbf{Y}|\mathbf{R}^N; \boldsymbol{\theta}_c, \boldsymbol{\theta}_v) = p^{\text{cond}}(\mathbf{Y}|\mathcal{M}(\mathbf{r}^n); \boldsymbol{\theta}_c, \boldsymbol{\theta}_v)$, and emphasize that the virtual particle distributions across both ensembles are dictated by the VCG model parameters. We now explicitly evaluate the VAA ensemble average in Equation 20 for a VCG model parameter involving virtual particles, $\theta_v$,

$$\left\langle \frac{\partial U_{\mathcal{G}}}{\partial \theta_v} \right\rangle_{\text{VAA}} = \iint \frac{\partial U_{\mathcal{G}}}{\partial \theta_v}(\mathcal{M}(\mathbf{r}^n), \mathbf{Y}) \, p_{\text{pre}}^{\text{ref}}(\mathbf{r}^n, \mathbf{Y}) \mathrm{d}\mathbf{r}^n \mathrm{d}\mathbf{Y} \qquad (24)$$

Substituting the conditional distribution as defined in Equation 23 produces

$$\left\langle \frac{\partial U_{\mathcal{G}}}{\partial \theta_v} \right\rangle_{\text{VAA}} = \int \left[ \int \frac{\partial U_{\mathcal{G}}}{\partial \theta_v}(\mathcal{M}(\mathbf{r}^n), \mathbf{Y}) p^{\text{cond}}(\mathbf{Y}|\mathcal{M}(\mathbf{r}^n); \boldsymbol{\theta}_c, \boldsymbol{\theta}_v) \mathrm{d}\mathbf{Y} \right] p_{\mathbf{r}}^{\text{FG}}(\mathbf{r}^n) \mathrm{d}\mathbf{r}^n \qquad (25)$$

The term in the brackets is the conditional expectation of the potential energy derivative, $\overline{\frac{\partial U_{\mathcal{G}}}{\partial \theta_v}}$. The VAA ensemble average can then be re-expressed as an AA ensemble average of a conditional expectation

$$\left\langle \frac{\partial U_{\mathcal{G}}}{\partial \theta_v} \right\rangle_{\text{VAA}} = \left\langle \overline{\frac{\partial U_{\mathcal{G}}}{\partial \theta_v}} \right\rangle_{\text{AA}} \tag{26}$$

Similarly, the VCG ensemble average term in Equation 20 can be re-expressed as a CG ensemble average and the relative entropy derivative can be expressed more symmetrically as

$$\frac{\partial S_{\text{rel}}}{\partial \theta_v} = \beta \left\langle \overline{\frac{\partial U_{\mathcal{G}}}{\partial \theta_v}} \right\rangle_{\text{AA}} - \beta \left\langle \overline{\frac{\partial U_{\mathcal{G}}}{\partial \theta_v}} \right\rangle_{\text{CG}} \tag{27}$$

Equation 27 is the central equation of this section, and demonstrates how virtual particle parameters can be optimized exactly within a relative entropy framework. We note that while the conditional expectation is evaluated in both reference and model ensembles in Equation 27, the conditional expectation itself is evaluated only over the model ensemble. This mixing of ensembles is not unexpected, as virtual particles are only "seen" in the model ensemble and thus all predictive modeling for virtual particles stems from the sampling of this ensemble.

In practice, the conditional expectation is not explicitly known. However, since the conditional expectation is the minimum of a least squares prediction, simulation of the VCG model can be used to obtain an approximation to the conditional expectation, $m_v(\mathbf{R}^N)$, through the following variational statement

$$\min_{m_v(\mathbf{R}^N)} \iint \left[ m_v(\mathbf{R}^N) - \frac{\partial U_{\mathcal{G}}}{\partial \theta_v}(\mathbf{R}^N, \mathbf{Y}) \right]^2 p_{\text{pre}}^{\text{mod}}(\mathbf{R}^N, \mathbf{Y}; \boldsymbol{\theta}_c, \boldsymbol{\theta}_v) d\mathbf{R}^N \, d\mathbf{Y}$$

$$= \overline{\frac{\partial U_{\mathcal{G}}}{\partial \theta_v}} \tag{28}$$

When an unlimited basis set to express the model interactions is utilized, the variational statement is exact. In practice, a limited basis set for the predictor is used, and an approximate predictor, $m_v{}^\dagger$, is obtained. The derivative of the relative entropy with respect to the considered virtual particle parameter can then be approximated by utilizing this predictor

$$\frac{\partial S_{\text{rel}}}{\partial \theta_v} \approx \beta \langle m_v{}^\dagger(\mathbf{R}^N) \rangle_{\text{AA}} - \beta \langle m_v{}^\dagger(\mathbf{R}^N) \rangle_{\text{CG}} \tag{29}$$

Virtual particle parameters can be optimized in practice by first solving the least squares regression problem presented in Equation 28 to develop an approximate model of the conditional expectation. The ensemble averages of this model are then taken with respect to the AA and CG ensembles to calculate the relative entropy gradient in Equation 29. We note that the regression problem presented for virtual particle optimization shares connections with representing macroscopic observables through CG variables.[48] In this sense, the averaging over and thus the removal of virtual particles can be seen as a kind of CG operation in and of itself.

## III. Implementation

Equations 28 and 29 suggest that a simple augmentation to a traditional REM algorithm can be implemented to include virtual particles within REM optimized CG models.[24] First, the VCG model is simulated for a given parameter set, and statistics pertinent to REM optimization are recorded. Second, for each virtual particle parameter to be optimized, a predictor is trained using real particle features to approximate the corresponding conditional averaged potential energy derivative through least-squares loss. Lastly, the ensemble averages of all potential energy derivatives are calculated in both the model and reference ensembles to calculate the gradient of the relative entropy, which is then used to update the parameters iteratively. This algorithm is summarized in Figure 2.

While the algorithm for virtual particle optimization is exact in principle, the approximate nature of the gradient descent along the virtual particle parameter space may affect optimization and model performance. Additionally, as only the first derivative of the virtual particle parameter in VD-REM is approximated, update schemes which incorporate higher-order derivatives cannot be utilized. We note the complexity of the predictor bears no additional computational cost on actual simulation of the VCG model, which features only pairwise interactions.

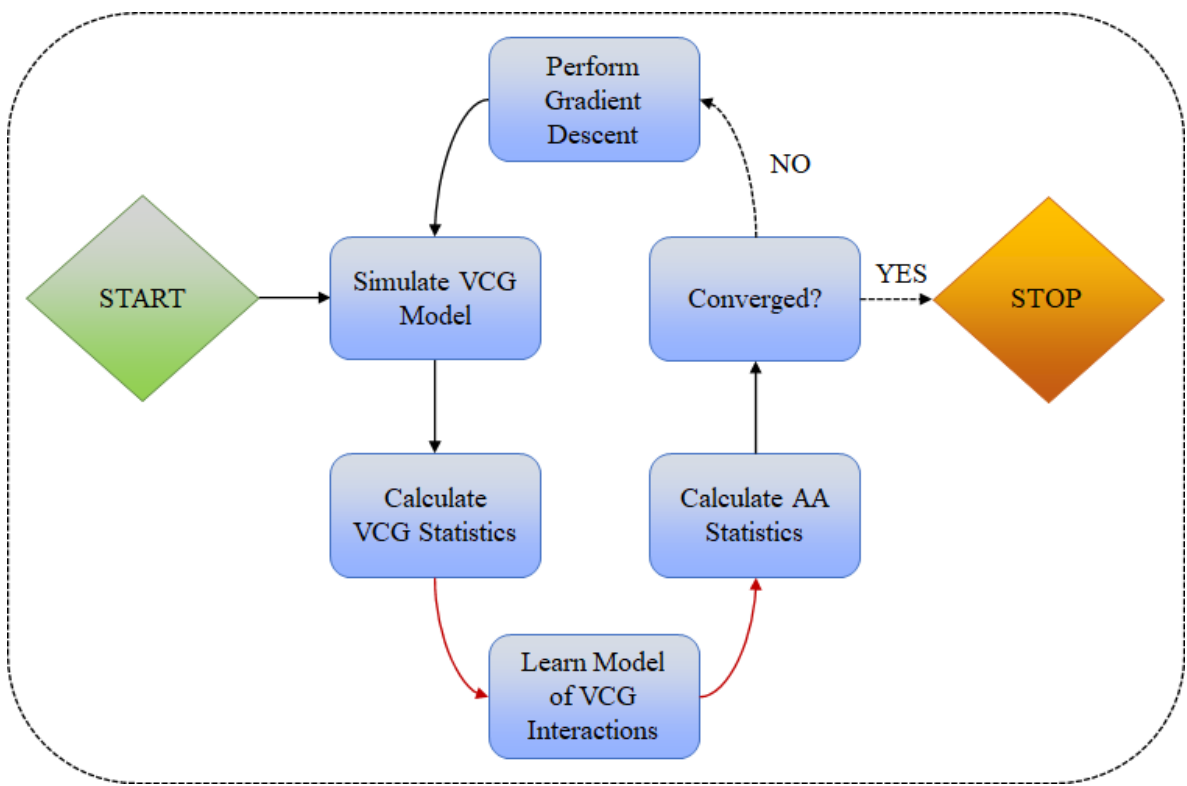

**Figure 2.** Workflow diagram of VD-REM. An initial model is fed into VD-REM which iteratively updates model parameters according to gradient descent. These steps are performed as in REM for real particles, while machine-learned models are learned to approximate virtual particle interactions first.

Furthermore, since the regression problem for each parameter is uniquely evaluated over the VCG ensemble, large amounts of data obtained through computational simulation can be acquired without rendering practical implementations infeasible. Consequently, we utilize machine learning (in particular, gradient boost models[49]) in constructing predictors for each potential energy derivative.

The features which comprise each descriptor must follow the symmetries present in the interactions it is intended to model. For the virtual particles considered here, these include translational, rotational, and permutation (between particles of the same type) symmetries. The features utilized in each descriptor consisted of pairwise nonbonded ($r$), bonded ($b$), and angular interactions ($a$). Two feature types were considered. For the first feature type, the set of all values in the considered interaction types (e.g., nonbonded pair) $x$ for a specific CG bead grouping $I$ (e.g, lipid head group and middle group) are selected for placement in a bin $B$ corresponding to a pre-selected range of values for that interaction type

$$F_1(x(I), B) = \frac{1}{|x(I)|} \sum_{i \in x(I)} 1 \cdot [i \in B] \tag{30}$$

The ranges considered for bin construction for each interaction type is detailed in the supporting information (SI). The second feature type utilized only pairwise interactions and consisted of pair-averaged moments for each CG interaction

$$F_2(r(I)) = \frac{1}{|r(I)|} \sum_{i \in r(I)} i^n \tag{31}$$

All pairwise interaction types were incorporated as features, and $n = 2, 4, 6$ and $12$ moments were considered.

Virtual particle interactions, including self-interactions and interactions with real CG beads, were implemented with a pre-defined model function, in this case, Lennard-Jones, to simplify training and aid in stability during simulation. For all CG bead types of type $I$ and $J$ the potential energy interaction can be defined as

$$U_{IJ}(\epsilon, \sigma) = \sum_{i \in I, j \in J} U_{IJ}(r_{ij}; \epsilon, \sigma) = 4\epsilon \sum_{i \in I, j \in J} \left[ \left(\frac{\sigma}{r_{ij}}\right)^{12} - \left(\frac{\sigma}{r_{ij}}\right)^{6} \right] \tag{32}$$

The virtual particle parameters to be optimized for this interaction are then $\theta_v = (\epsilon, \sigma)$ Consequently, for each virtual particle interaction considered a predictor is constructed for the conditional expectation of the potential energy derivatives, $\overline{\frac{\partial U_{IJ}}{\partial \epsilon}}$ and $\overline{\frac{\partial U_{IJ}}{\partial \sigma}}$. These predictors are then used to approximate the reference ensemble average used during gradient descent.

In practice, when a limited basis is used for the predictor, the generalization of the predictor across ensembles may be poor when the reference and model ensembles strongly differ. A similar issue occurs for virtual particle optimization in ARCG,[46] and in both cases is most likely to occur at the beginning of training. While the exact predictor will generalize completely across CG and AA ensembles, model predictors must learn only from the CG ensemble and will inherently only be able to generalize to the AA ensemble in a limited capacity. Additionally, previous work on REM has established the gradient of the relative entropy to be noisy;[50] we note it was found during training of the gradient boost models that substantial regularization was required to avoid overfitting. This is related to the insufficient sampling of the CG ensemble inherent to REM, which has necessitated additional regularization strategies during gradient descent. We suggest two alternative routes to circumvent these issues: (1) defer updating virtual particle interactions from their initialized values while VD-REM iterations exhibit strong disparities between the reference and model ensembles and (2) initialize parameters in the VCG model to reduce disparities between the two ensembles. The latter route, which we pursue in the DOPC example, can be accomplished by utilizing a REM-optimized CG model as the starting point for training and a model-specific initial guess for the virtual particle interactions.

**IV. Proof of Concept**

Lipids are generally characterized by a hydrophilic head region and hydrophobic tail region. The amphipathic nature of lipids produces thermodynamically stable aggregates such as fluid bilayers, which serve as the template for cellular membranes.[51] CG models of lipid bilayers are highly desirable due to the large amount of solvent require for AA bilayer simulations, and the inherently slow lateral mixing of lipids.[52] The removal of solvent (i.e., a solvent-free model) necessitates encapsulation of the hydrophobic effect, a many-body interaction, within the remaining CG lipid beads.[53, 54]

While various bottom-up CG models of lipid bilayers have been developed,[42, 55-60] bottom-up CG methods have generally been unable to replicate the self-assembling features of lipids from an initial randomized mixture.

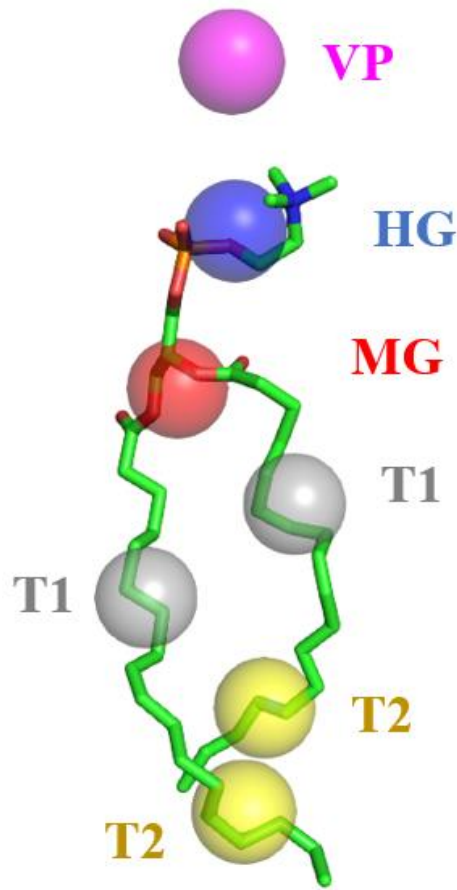

**Figure 3.** VCG mapping scheme for DOPC. Hydrogens are omitted for clarity. Each DOPC lipid is mapped to six real sites of four types: HG (blue), MG (red), T1 (gray), and T2 (yellow). A virtual particle, VP (purple), with no atomistic correspondence is then appended to the HG bead.

This suggests the bilayer conformation of solvent-free bottom-up CG models may represent only a metastable state, and that the hydrophobic effect cannot be represented without some sort of implicit solvent representation. The only bottom-up CG model which, to our knowledge, is capable of self-assembly utilizes a semi-explicit representation of solvent, in which the interfacial water of the hydrophilic region is appended as a "virtual particle" to the head group bead through a non-linear mapping.[42] The (highly) CG DOPC model consists of a head group (HG) bead representing the phosphocholine moiety, a middle group (MG) bead representing the glycerol center and ester group, outermost tail bead (T1) and innermost tail bead (T2). Inspired by this approach, we utilize a similar topology to construct a 7-site VCG model optimized using VD-REM, we denote VD-REM-7, in which a virtual particle (VP) with no atomistic correspondence is placed near the HG. This CG topology is represented in Figure 3.

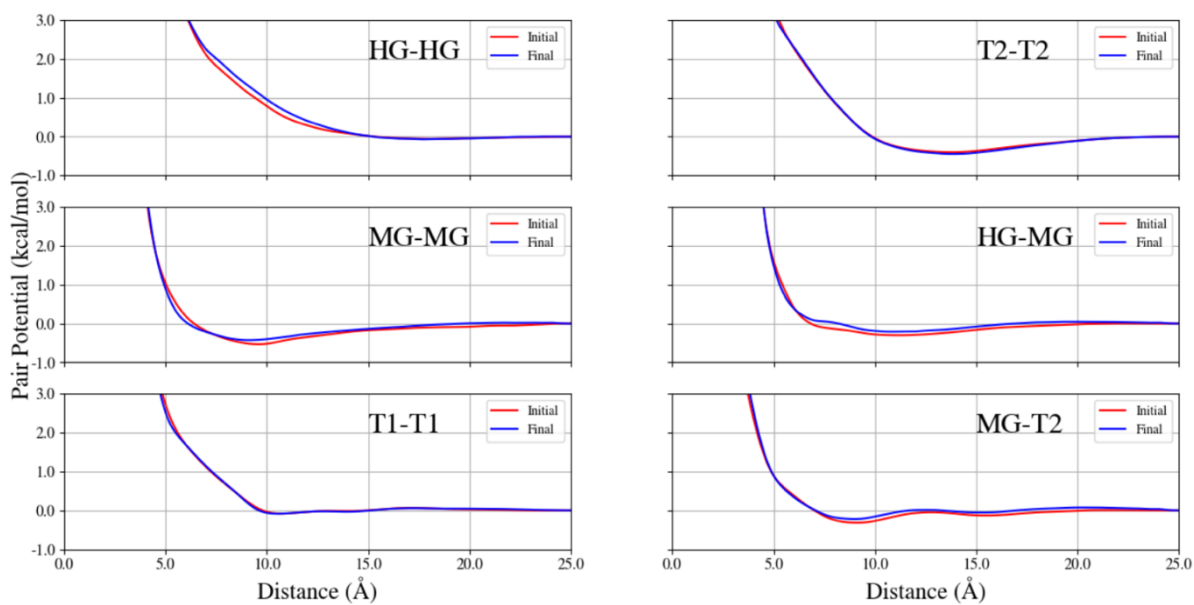

**Figure 4.** Subset of pair potentials for real CG particles at beginning and end of training. Initial potentials (red) which were adapted from the REM-6 model[42] and final VD-REM-7 model potentials (red) are shown.

**All-Atom Molecular Dynamics.** Reference simulations of a solvated DOPC bilayer were taken from Ref.[42] Briefly, 1152 DOPC lipids, ~ 45000 water molecules and 0.15 M NaCl were simulated using GROMACS 5.0.7 for 100 ns in the constant NVT ensemble at a temperature of 300 K as maintained by a Nose-Hoover thermostat.[61, 62] Simulations were initialized from a membrane area-equilibrated structure according to established protocol.[63, 64] The CHARMM36 force field was used for lipids and the TIP3P force field was used for water.[43, 65]

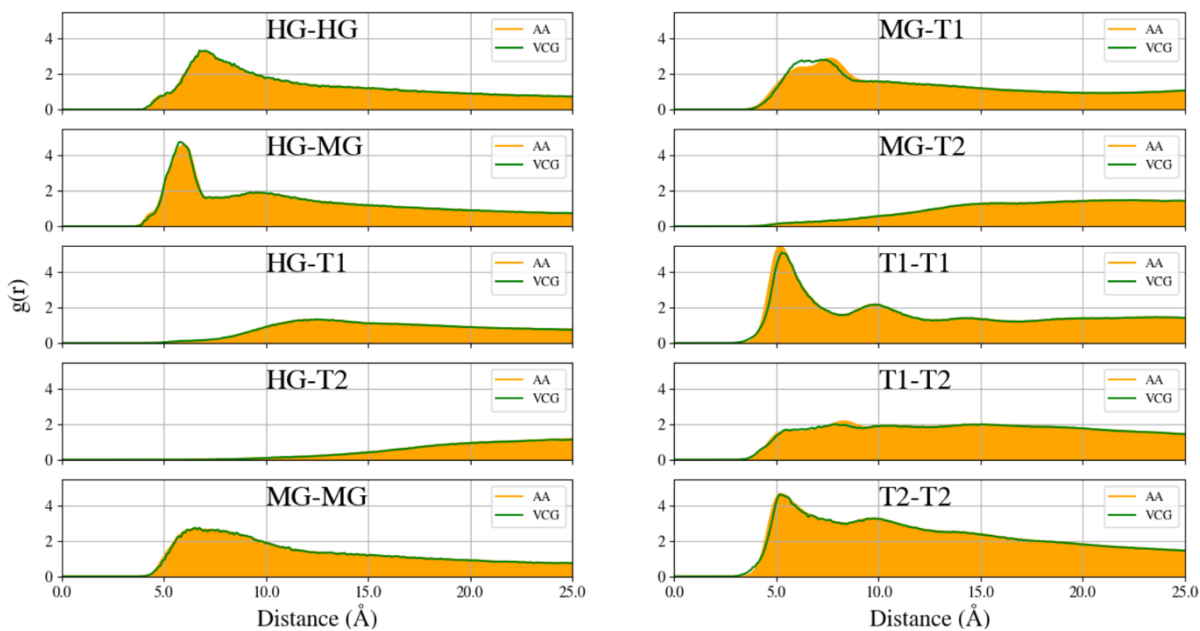

**Figure 5.** Radial distribution functions for real CG beads of DOPC. Both reference AA statistics (orange) and VCG statistics (green) are plotted.

**VD-REM.** Each pair interaction between real particles was described using 4$^{th}$ order B-splines with knots placed every 0.5 Å. Throughout training, the first B-spline knot value was held fixed, and the last three-B-splines were held fixed to maintain a zero value at the interaction cutoff. Initial interactions for real particles came from the REM-optimized model from Ref.[42] The mass of the virtual particle was set to 100 g/mol. Bonded interactions for the virtual particle to the CG lipid were represented as harmonic with an equilibrium constant of 1.5 kcal/mol Å and equilibrium distance of 2.5 Å. Angular virtual particle interactions were also considered harmonic with an equilibrium constant of 2.45 kcal/mol and equilibrium angle of 135°. Only nonbonded interactions were optimized during training. All virtual particle nonbonded interactions were initialized with LJ parameters $\epsilon = 0.05$ kcal/mol for all interactions and $\sigma = 4.0, 4.0, 4.5$ Å for VP-VP, HG-VP, and MG-VP interactions, respectively. Due to inherently minimal sampling between the hydrophobic and hydrophilic region, virtual particles and tail groups were considered non-interacting during training. All nonbonded interactions utilized a 25 Å cutoff. To improve stability during training, a maximum cap of 0.002 kcal/mol was permitted for B-spline coefficient updates, 0.0005 kcal/mol for $\epsilon$ updates, and 0.005 Å for $\sigma$ updates for the first 100 iterations. A more refined search was then conducted for 45 iterations with caps of 0.001 kcal/mol permitted for B-spline coefficient updates, 0.0001 kcal/mol for $\epsilon$ updates, and 0.001 Å for $\sigma$ updates to obtain the final VD-REM-7 model. For each virtual particle interaction parameter, LightGBM was used to construct a gradient boost model to predict the conditional potential energy derivative.[49] For each gradient boost model, the number of leaves was set to 5, the learning rate was set to 0.05, and 1000 estimators were chosen. The feature fraction and bagging fraction were set to 0.8 and 0.15, respectively. The L2 regularization hyperparameter was set to 20, and each tree was set to a max depth of 5. All other hyperparameters were set to default values unless specified. Information on feature construction for the machine-learned models is described in the SI.

**CG Molecular Dynamics.** At each VD-REM iteration, the following simulation protocol was conducted: The initial configuration for each simulation consisted of a bilayer-like lattice configuration with each lipid spaced 2 nm apart along the xy-plane. The initial structure was then minimized and simulated in the constant NPT ensemble for 25000 steps with a Langevin thermostat[66] and Berendsen barostat[67] with 2 ps and 5 ps damping constants, respectively. Linear deformation of the resulting structure to the reference atomistic lateral dimensions was then conducted for 25000 timesteps. A constant NVT simulation was then run for 1200000 timesteps in which 50000 frames were collected as training data for least-squares regression and to record real particle ensemble averages. A timestep of 5.0 fs was used throughout training. Large-scale production simulations of the resulting VD-REM model were performed in the constant NVT ensemble with 10,368 lipids and a 10 fs timestep to collect statistics and spectral information. All CG MD simulations were conducted using the LAMMPS MD engine.[68]

**Results.** A subset of real particle potentials are shown in Figure 4. It is apparent from Figure 4 that the VD-REM alters the real particle interactions mainly by weakening attractive wells, as seen in the MG-MG, HG-MG, and MG-T2 interactions. Tail interactions are augmented to a lesser degree, which is likely explained by the absence of interactions with VPs. Virtual particle interactions of the optimized VD-REM-7 model consisted of LJ parameters $\epsilon = 0.0328, 0.0052,$ and $0.0210$ kcal/mol and $\sigma = 3.97, 3.59,$ and $3.98$ Å for VP-VP, HG-VP, and MG-VP interactions, respectively. Consequently, the resulting interactions of the virtual particles in the VD-REM-7 model predominantly consist of hard wall interactions with minimal attraction.

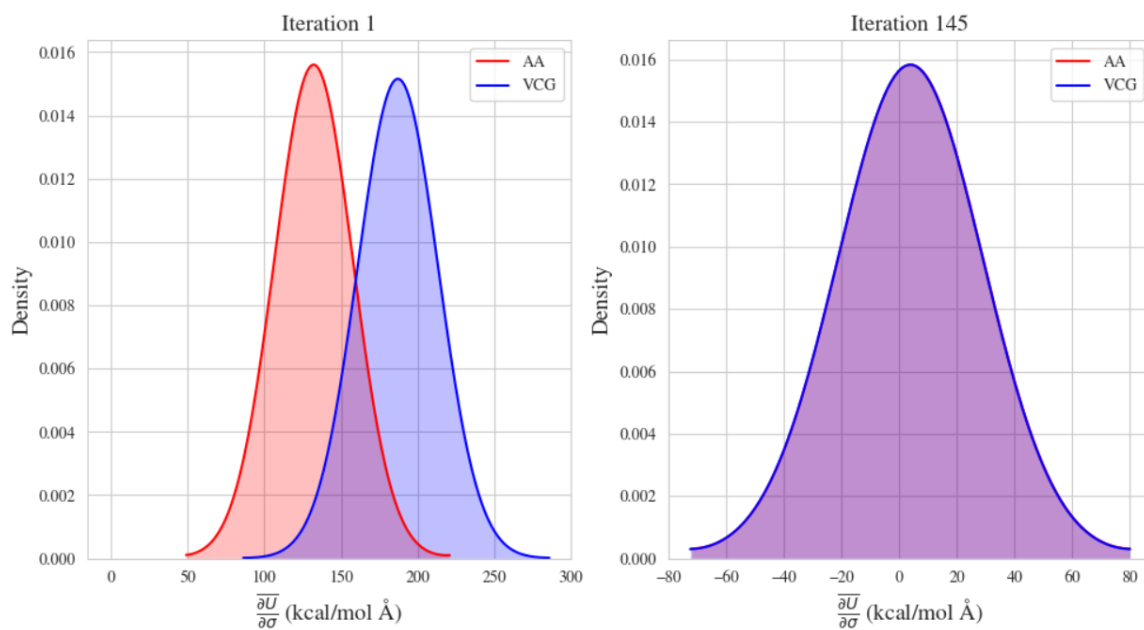

**Figure 6.** Predicted values for the HG-VP potential energy derivative $\overline{\frac{\partial U}{\partial \sigma}}$ across VCG and AA ensembles. The predicted values for the initial (left) and final (right) iterations are plotted using kernel density estimation with a bin width of 25 kcal/mol Å.

Three-dimensional radial distribution functions (RDF) for all real particles are shown in Figure 5. As expected, the fidelity of pair correlations in REM is also found in VD-REM. In principle, the optimization of parameters in the LJ potential describing interactions of the VPs should guarantee recapitulation of $r^{-12}$ and $r^{-6}$ statistics. However, the absence of an explicit representation of virtual particles in the reference ensemble precludes this direct analysis.

Instead, we analyze the convergence of the ensemble averages of the pertinent potential energy derivatives in Equation 27 between AA and VCG ensembles, as in Figure 6. The machine-learned model constructed for each potential energy derivative was used to predict framewise values across both ensembles.

**Table 1.** Material properties of various models of DOPC including bilayer thickness ($d_{\text{HG-HG}}$), orientational order parameter ($S_D$), and bending modulus ($\kappa$).

| Model | $d_{\text{HG-HG}}$ (Å) | $S_D$ | $\kappa$ ($k_B T$) |
|---|---|---|---|
| Exp.[69] | 4.48 | n/a | 18.3 |
| AA[42] | 4.0 | 0.56 | 23.4 ± 2.6,28[70] |
| MSCG-6[42] | 4.4 | 0.65 | 56.4 ± 2.2 |
| REM-6[42] | 4.1 | 0.63 | 183.9 ± 10.4 |
| VD-REM-7 | 4.1 | 0.63 | 81.6 ± 3.1 |

Analysis of material properties is presented in Table 1. Bilayer thickness was calculated from the bin-averaged normal distance between HG beads of each leaflet using a bin size of 1.0 nm. Orientational order parameters were calculated from the second-order Legendre polynomial for $\cos \theta$, such that[71]

$$S_D(\theta) = \left\langle \frac{1}{2}(3\cos^2(\theta) - 1) \right\rangle \tag{33}$$

in which $\theta$ is the angle the vector each bonded T1 and T2 makes with the bilayer normal.

The bending modulus, $\kappa$, was calculated via connections to Canham-Helfrich continuum theory.[72, 73] In the tensionless ensemble, the height fluctuation spectrum $u$ with associated reciprocal space vector $\mathbf{q}$ is connected to the bending modulus through the following relation

$$\langle A|u(q)|^2 \rangle = \frac{k_B T}{\kappa q^4} \tag{34}$$

where *A* is the instantaneous bilayer area. This method has been previously used to calculate the bending modulus for AA and CG bilayer simulations.[12,42,74,75] Membrane rigidity is a key biophysical metric in describing membrane-protein interactions.[76,77] This fitting was performed using data from a large-scale production run to obtain the bending modulus of VD-REM-7 DOPC (see Figure 7).

We note that, according to Table 1, properties such as bilayer thickness and orientational order parameter stayed consistent between REM-6 and VD-REM-7. However, the bending modulus of the VD-REM is less than half of the much too stiff REM DOPC model, indicating that the VD-REM is significantly more flexible. This suggests the introduction of virtual particles into the model has enabled recapitulation, to some extent, of higher order correlations. We note that the MS-CG-6 model of DOPC retains the most accurate bending modulus of the CG models presented. This is likely due to the MS-CG-6 model's explicit attempt to relate two- and three-body correlations via the MS-CG algorithm. While the virtual particle methodology presented here provides no method of optimization outside of REM, we detail in the Discussion section possible routes to work around this.

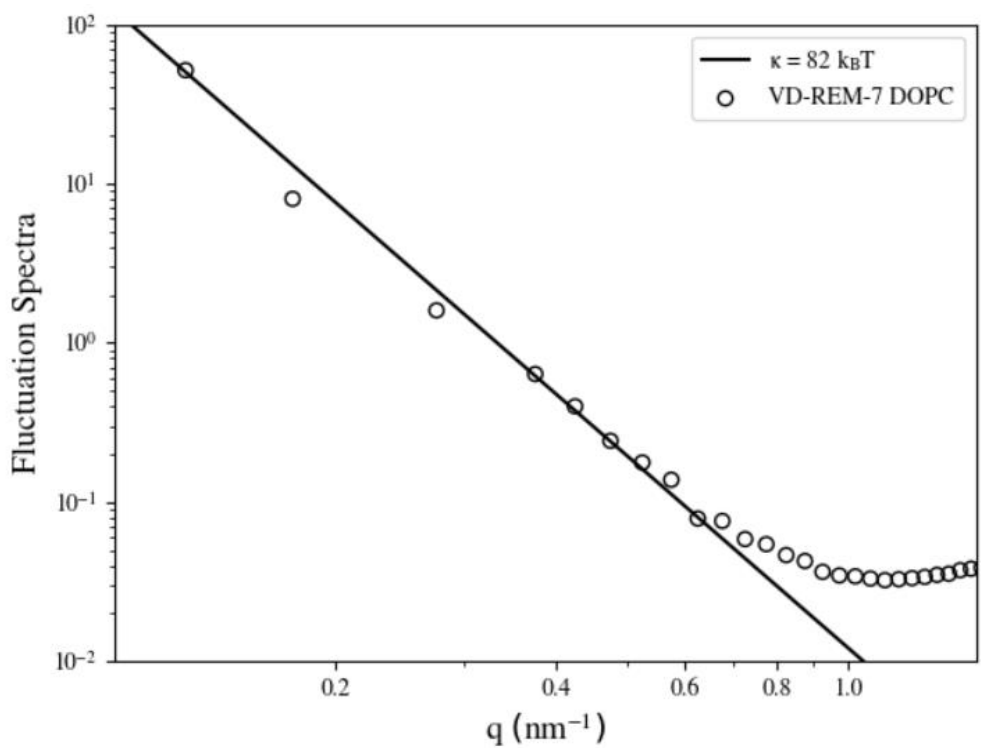

**Figure 7.** Height fluctuation spectra of VD-REM-7 DOPC as a function of wavenumber. Simulation data (circles) were fit according to Equation 34 (solid line) to obtain the bending modulus.

As demonstrated in Figure 8 and Movie S1, the VD-REM-7 DOPC model is also capable of self-assembly starting from a random configuration. This suggests the hydrophobic effect is more adequately represented in the VD-REM-7 model than the REM-6 model, which is unable to properly self-assemble. Since the VD-REM-7 DOPC model is entirely anhydrous, the hydrophobic effect is interpreted within model interactions implicitly through the reference statistics of the CG lipids.

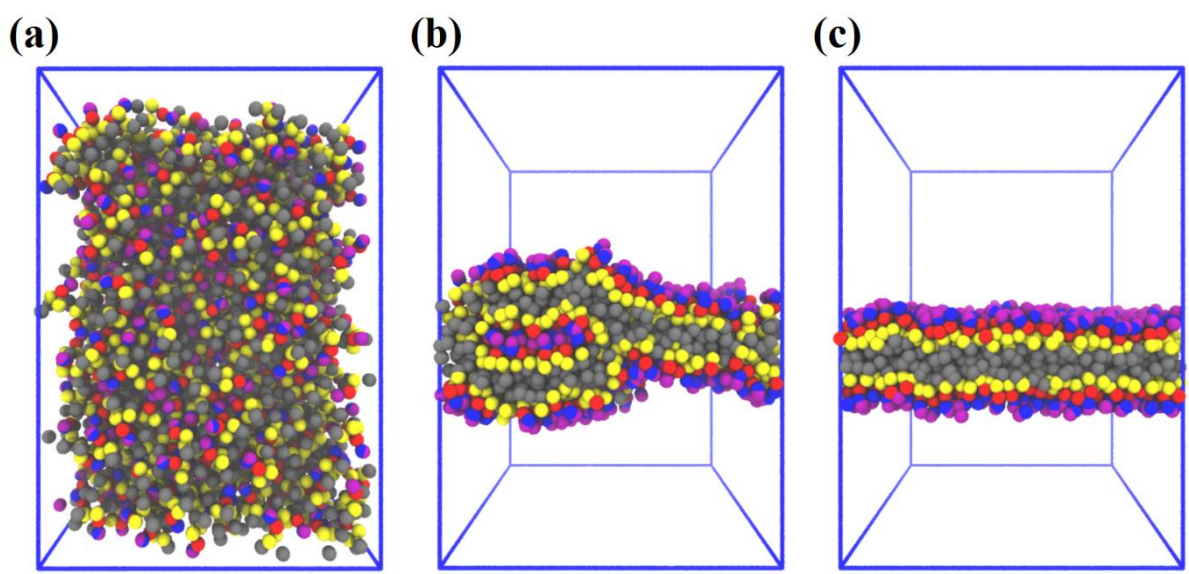

**Figure 8.** Depiction of the VD-REM-7 DOPC model at initial (a), intermediate (b), and final (c) stages of self assembly.

Although we must be careful to describe the self-assembling nature of the VD-REM-7 model holistically, we note that self-assembly is lost upon rendering the virtual particle non-interacting, as shown in Figure S1. The location of the virtual particles in the context of the bilayer configuration, as well as the predominantly repulsive nature of their interactions, suggests their addition into the model may introduce a general cohesive force which maintains bilayer stability.

## V. Discussion

We note that the methodology presented here, in which optimization of a CG model is decoupled from the resolution during simulation, bears conceptual similarities to predictive coarse-graining (PCG).[47] In PCG, a nondeterministic backmap from the CG to AA resolution is constructed via expectation-maximization to critique the underlying FG statistics generated by the CG model. The approach for virtual particle optimization described here has been entirely discriminative, i.e., focused on the conditional expectation of the target functions. Instead, a generative approach similar to PCG could be taken to construct an analogous CG to VCG backmapping operator. Optimization of the virtual particle interaction parameters would then proceed exactly as in CG REM, but ensemble averages of target functions are evaluated over an approximate joint distribution of virtual and real particle configurational degrees of freedom. Furthermore, this generative approach would allow for the establishment of local information of virtual particles in the reference ensemble through the joint distribution, enabling optimization in force-based methods such as MS-CG and generalized Yvon-Born-Green theory.[13-16, 78] We intend to further establish this connection in future work.

Since the introduction of virtual particles into CG models is not restricted by an AA to VCG mapping, the systematic optimization presents new avenues by which virtual particles can augment CG models. Given the critical role solvent plays in protein behavior,[79, 80] highly CG models of large proteins such as actin-related proteins,[81] HIV-1 capsid (CA) protein, and BAR domain proteins[82] may benefit from introduction of virtual particles.[40, 83] Alternatively, it has been demonstrated complex lipid behavior including gel-phase hexatic tail ordering and the structural diversity of the ripple phase are not fully recapitulated in low-resolution CG models.[42, 84] One could in principle introduce virtual particles to lipid tails to attempt to remediate this.

Conversely, the lack of a reference mapping for virtual particles renders their meaning within the context of CG modeling somewhat nebulous. The representation of virtual particle behavior in the reference ensemble is inherently approximate and is represented via a machine-learned function of the real particle statistics. It is worth considering how the features inputted into these models relate to the function variationally obtained, i.e., what features of the real particle statistics are considered important when the model is learned. Explainable machine learning, in particular Shapley additive explanations,[85] could be utilized to explore this relation.

Lastly, we note that alternative machine learning approaches could be utilized in solving the regression problems presented in VD-REM beyond gradient boost models. The least-squares regression problem presented in Equation 28 relates the real particle configurational information to a scalar free energy-like function of the virtual particles. This problem is conceptually similar to other machine-learning approaches to CG free energy surfaces,[23, 32, 33] as well as the fitting of atomic data to potential energy surfaces.[86-91] Consequently, it may be fruitful to pursue similar machine learning approaches which incorporate neural networks and Gaussian processes in construction of model predictors.

## VI. Conclusions

In this work, we have introduced a new coarse-graining methodology, variational derivative relative entropy minimization (VD-REM), to systematically introduce virtual particles, having no explicit atomistic correspondence, within a relative entropy framework. We demonstrate how virtual particles can be optimized in the absence of their reference statistics through machine-learning models of pertinent target functions. We apply this methodology to construct a solvent-free, low-resolution CG model of a 1,2-dioleoyl-sn-glycero-3-phosphocholine (DOPC) lipid bilayer and demonstrate that introduction of virtual particles about the lipid heads enables self-assembly within a CG model that is otherwise incapable of self-assembly. We additionally show that higher order correlations are matched more effectively utilizing virtual particles through a substantial reduction of the CG bending modulus. These results suggest virtual particles will be fruitful in recapitulating solvent-mediated behavior, as well as higher order correlations, within implicit solvent, highly CG models. Although our proof-of-concept demonstrates the utility of virtual particles, a substantial analysis is warranted into alternate machine-learning models such as Gaussian processes and neural networks which may prove more beneficial in predicting the target functions unique to VD-REM. Additionally, feature importance techniques may be informative on representing more explicitly how virtual particles may improve CG models.

Taken as a whole, the directions outlined in this work promise to lead to entirely new, more flexible, and more accurate bottom-up CG models, along with their subsequent modeling of real systems at the CG level.

**AUTHOR INFORMATION**

**Corresponding Author**

Email: gavoth@uchicago.edu

ORCID

Gregory A. Voth: 0000-0002-3267-6748

**ACKNOWLEDGEMENTS**

This material is based upon work supported in part by the National Science Foundation (NSF grant CHE-2102677) and in part by the National Institutes of Health (NIH grant R01GM063796). Simulations were performed using computing resources provided by the University of Chicago Research Computing Center (RCC). T.D.L. was also supported by the National Science Foundation Graduate Research Fellowship (DGE-1746045).

# REFERENCES

1.	Brooks, C. L., Computer simulation of liquids. *J. Solution Chem.* **1989,** *18* (1), 99-99.
2.	Karplus, M.; McCammon, J. A., Molecular dynamics simulations of biomolecules. *Nat Struct Biol* **2002,** *9* (9), 646-52.
3.	Dror, R. O.; Dirks, R. M.; Grossman, J. P.; Xu, H.; Shaw, D. E., Biomolecular simulation: a computational microscope for molecular biology. *Annu Rev Biophys* **2012,** *41*, 429-52.
4.	Pak, A. J.; Voth, G. A., Advances in coarse-grained modeling of macromolecular complexes. *Curr. Opin. Struct. Biol.* **2018,** *52*, 119-126.
5.	Tozzini, V., Coarse-grained models for proteins. *Curr Opin Struct Biol* **2005,** *15* (2), 144-50.
6.	Voth, G. A., *Coarse-Graining of Condensed Phase and Biomolecular Systems*. CRC Press: 2009.
7.	Noid, W. G., Perspective: Coarse-grained models for biomolecular systems. *J Chem Phys* **2013,** *139* (9), 090901.
8.	Jin, J.; Pak, A. J.; Durumeric, A. E. P.; Loose, T. D.; Voth, G. A., Bottom-up Coarse-Graining: Principles and Perspectives. *J Chem Theory Comput* **2022,** *18* (10), 5759-5791.
9.	Monticelli, L.; Kandasamy, S. K.; Periole, X.; Larson, R. G.; Tieleman, D. P.; Marrink, S. J., The MARTINI Coarse-Grained Force Field: Extension to Proteins. *J Chem Theory Comput* **2008,** *4* (5), 819-34.
10.	Risselada, H. J.; Marrink, S. J., The molecular face of lipid rafts in model membranes. *Proc Natl Acad Sci U S A* **2008,** *105* (45), 17367-72.
11.	Souza, P. C. T.; Alessandri, R.; Barnoud, J.; Thallmair, S.; Faustino, I.; Grunewald, F.; Patmanidis, I.; Abdizadeh, H.; Bruininks, B. M. H.; Wassenaar, T. A.; Kroon, P. C.; Melcr, J.; Nieto, V.; Corradi, V.; Khan, H. M.; Domanski, J.; Javanainen, M.; Martinez-Seara, H.; Reuter, N.; Best, R. B.; Vattulainen, I.; Monticelli, L.; Periole, X.; Tieleman, D. P.; de Vries, A. H.; Marrink, S. J., Martini 3: a general purpose force field for coarse-grained molecular dynamics. *Nat Methods* **2021,** *18* (4), 382-388.
12.	Jarin, Z.; Newhouse, J.; Voth, G. A., Coarse-Grained Force Fields from the Perspective of Statistical Mechanics: Better Understanding of the Origins of a MARTINI Hangover. *J Chem Theory Comput* **2021,** *17* (2), 1170-1180.
13.	Izvekov, S.; Voth, G. A., A multiscale coarse-graining method for biomolecular systems. *J Phys Chem B* **2005,** *109* (7), 2469-73.
14.	Noid, W. G.; Chu, J. W.; Ayton, G. S.; Voth, G. A., Multiscale coarse-graining and structural correlations: connections to liquid-state theory. *J Phys Chem B* **2007,** *111* (16), 4116-27.
15.	Noid, W. G.; Chu, J. W.; Ayton, G. S.; Krishna, V.; Izvekov, S.; Voth, G. A.; Das, A.; Andersen, H. C., The multiscale coarse-graining method. I. A rigorous bridge between atomistic and coarse-grained models. *J Chem Phys* **2008,** *128* (24), 244114.
16.	Noid, W. G.; Liu, P.; Wang, Y.; Chu, J. W.; Ayton, G. S.; Izvekov, S.; Andersen, H. C.; Voth, G. A., The multiscale coarse-graining method. II. Numerical implementation for coarse-grained molecular models. *J Chem Phys* **2008,** *128* (24), 244115.
17.	Shell, M. S., The relative entropy is fundamental to multiscale and inverse thermodynamic problems. *The Journal of Chemical Physics* **2008,** *129* (14), 144108.
18.	Chaimovich, A.; Shell, M. S., Relative entropy as a universal metric for multiscale errors. *Physical Review E* **2010,** *81* (6), 060104.

19. Dunn, N. J.; Foley, T. T.; Noid, W. G., Van der Waals Perspective on Coarse-Graining: Progress toward Solving Representability and Transferability Problems. *Acc Chem Res* **2016,** *49* (12), 2832-2840.
20. Wagner, J. W.; Dama, J. F.; Durumeric, A. E.; Voth, G. A., On the representability problem and the physical meaning of coarse-grained models. *J Chem Phys* **2016,** *145* (4), 044108.
21. Jin, J.; Yu, A.; Voth, G. A., Temperature and Phase Transferable Bottom-up Coarse-Grained Models. *J Chem Theory Comput* **2020,** *16* (11), 6823-6842.
22. Wang, J.; Charron, N.; Husic, B.; Olsson, S.; Noe, F.; Clementi, C., Multi-body effects in a coarse-grained protein force field. *J. Chem. Phys.* **2021,** *154* (16), 164113.
23. John, S. T.; Csanyi, G., Many-Body Coarse-Grained Interactions Using Gaussian Approximation Potentials. *J Phys Chem B* **2017,** *121* (48), 10934-10949.
24. Chaimovich, A.; Shell, M. S., Coarse-graining errors and numerical optimization using a relative entropy framework. *J Chem Phys* **2011,** *134* (9), 094112.
25. Larini, L.; Lu, L.; Voth, G. A., The multiscale coarse-graining method. VI. Implementation of three-body coarse-grained potentials. *J Chem Phys* **2010,** *132* (16), 164107.
26. Wagner, J. W.; Dannenhoffer-Lafage, T.; Jin, J.; Voth, G. A., Extending the range and physical accuracy of coarse-grained models: Order parameter dependent interactions. *J Chem Phys* **2017,** *147* (4), 044113.
27. Dama, J. F.; Jin, J.; Voth, G. A., The Theory of Ultra-Coarse-Graining. 3. Coarse-Grained Sites with Rapid Local Equilibrium of Internal States. *J Chem Theory Comput* **2017,** *13* (3), 1010-1022.
28. Sanyal, T.; Shell, M. S., Coarse-grained models using local-density potentials optimized with the relative entropy: Application to implicit solvation. *J Chem Phys* **2016,** *145* (3), 034109.
29. Sanyal, T.; Shell, M. S., Transferable Coarse-Grained Models of Liquid-Liquid Equilibrium Using Local Density Potentials Optimized with the Relative Entropy. *J Phys Chem B* **2018,** *122* (21), 5678-5693.
30. Lafond, P. G.; Izvekov, S., Multiscale Coarse-Graining of Polarizable Models through Force-Matched Dipole Fluctuations. *J Chem Theory Comput* **2016,** *12* (12), 5737-5750.
31. Lemke, T.; Peter, C., Neural Network Based Prediction of Conformational Free Energies - A New Route toward Coarse-Grained Simulation Models. *J Chem Theory Comput* **2017,** *13* (12), 6213-6221.
32. Zhang, L.; Han, J.; Wang, H.; Car, R.; E, W., DeePCG: Constructing coarse-grained models via deep neural networks. *J Chem Phys* **2018,** *149* (3), 034101.
33. Wang, J.; Olsson, S.; Wehmeyer, C.; Perez, A.; Charron, N. E.; de Fabritiis, G.; Noe, F.; Clementi, C., Machine Learning of Coarse-Grained Molecular Dynamics Force Fields. *ACS Cent Sci* **2019,** *5* (5), 755-767.
34. Wang, W.; Gómez-Bombarelli, R., Coarse-graining auto-encoders for molecular dynamics. *npj Computational Materials* **2019,** *5* (1), 125.
35. Husic, B. E.; Charron, N. E.; Lemm, D.; Wang, J.; Perez, A.; Majewski, M.; Kramer, A.; Chen, Y.; Olsson, S.; de Fabritiis, G.; Noe, F.; Clementi, C., Coarse graining molecular dynamics with graph neural networks. *J Chem Phys* **2020,** *153* (19), 194101.
36. Reith, D.; Putz, M.; Muller-Plathe, F., Deriving effective mesoscale potentials from atomistic simulations. *J. Comput. Chem.* **2003,** *24* (13), 1624-36.
37. Lyubartsev, A. P.; Laaksonen, A., Calculation of effective interaction potentials from radial distribution functions: A reverse Monte Carlo approach. *Phys. Rev. E* **1995,** *52* (4), 3730-3737.

38. Rudzinski, J. F.; Noid, W. G., Coarse-graining entropy, forces, and structures. *J Chem Phys* **2011,** *135* (21), 214101.
39. Jin, J.; Han, Y.; Pak, A. J.; Voth, G. A., A new one-site coarse-grained model for water: Bottom-up many-body projected water (BUMPer). I. General theory and model. *J Chem Phys* **2021,** *154* (4), 044104.
40. Grime, J. M. A.; Dama, J. F.; Ganser-Pornillos, B. K.; Woodward, C. L.; Jensen, G. J.; Yeager, M.; Voth, G. A., Coarse-grained simulation reveals key features of HIV-1 capsid self-assembly. *Nat Commun* **2016,** *7*, 11568.
41. Jin, J.; Han, Y.; Voth, G. A., Coarse-graining involving virtual sites: Centers of symmetry coarse-graining. *J Chem Phys* **2019,** *150* (15), 154103.
42. Pak, A. J.; Dannenhoffer-Lafage, T.; Madsen, J. J.; Voth, G. A., Systematic Coarse-Grained Lipid Force Fields with Semiexplicit Solvation via Virtual Sites. *J Chem Theory Comput* **2019,** *15* (3), 2087-2100.
43. Jorgensen, W. L.; Chandrasekhar, J.; Madura, J. D.; Impey, R. W.; Klein, M. L., Comparison of simple potential functions for simulating liquid water. *The Journal of Chemical Physics* **1983,** *79* (2), 926-935.
44. Melo, M. N.; Ingolfsson, H. I.; Marrink, S. J., Parameters for Martini sterols and hopanoids based on a virtual-site description. *J Chem Phys* **2015,** *143* (24), 243152.
45. Liu, Y.; De Vries, A. H.; Barnoud, J.; Pezeshkian, W.; Melcr, J.; Marrink, S. J., Dual Resolution Membrane Simulations Using Virtual Sites. *J Phys Chem B* **2020,** *124* (19), 3944-3953.
46. Durumeric, A. E. P.; Voth, G. A., Adversarial-residual-coarse-graining: Applying machine learning theory to systematic molecular coarse-graining. *J Chem Phys* **2019,** *151* (12), 124110.
47. Schöberl, M.; Zabaras, N.; Koutsourelakis, P.-S., Predictive coarse-graining. *Journal of Computational Physics* **2017,** *333*, 49-77.
48. Dannenhoffer-Lafage, T.; Wagner, J. W.; Durumeric, A. E. P.; Voth, G. A., Compatible observable decompositions for coarse-grained representations of real molecular systems. *J Chem Phys* **2019,** *151* (13), 134115.
49. Ke, G.; Meng, Q.; Finley, T.; Wang, T.; Chen, W.; Ma, W.; Ye, Q.; Liu, T.-Y., LightGBM: a highly efficient gradient boosting decision tree. In *Proceedings of the 31st International Conference on Neural Information Processing Systems*, Curran Associates Inc.: Long Beach, California, USA, 2017; pp 3149–3157.
50. Bilionis, I.; Zabaras, N., A stochastic optimization approach to coarse-graining using a relative-entropy framework. *J Chem Phys* **2013,** *138* (4), 044313.
51. Simons, K.; Sampaio, J. L., Membrane organization and lipid rafts. *Cold Spring Harb. Perspect. Biol.* **2011,** *3* (10), a004697.
52. Hong, C.; Tieleman, D. P.; Wang, Y., Microsecond molecular dynamics simulations of lipid mixing. *Langmuir* **2014,** *30* (40), 11993-2001.
53. Chandler, D., Interfaces and the driving force of hydrophobic assembly. *Nature* **2005,** *437* (7059), 640-7.
54. Marsh, D., Equation of State for Phospholipid Self-Assembly. *Biophys J* **2016,** *110* (1), 188-96.
55. Izvekov, S.; Voth, G. A., Solvent-free lipid bilayer model using multiscale coarse-graining. *J Phys Chem B* **2009,** *113* (13), 4443-55.
56. Lu, L.; Voth, G. A., Systematic coarse-graining of a multicomponent lipid bilayer. *J Phys Chem B* **2009,** *113* (5), 1501-10.
57. Wang, Z. J.; Deserno, M., A systematically coarse-grained solvent-free model for quantitative phospholipid bilayer simulations. *J Phys Chem B* **2010,** *114* (34), 11207-20.

58. Sodt, A. J.; Head-Gordon, T., An implicit solvent coarse-grained lipid model with correct stress profile. *J Chem Phys* **2010,** *132* (20), 205103.
59. Srivastava, A.; Voth, G. A., A Hybrid Approach for Highly Coarse-grained Lipid Bilayer Models. *J Chem Theory Comput* **2013,** *9* (1), 750-765.
60. Srivastava, A.; Voth, G. A., Solvent-Free, Highly Coarse-Grained Models for Charged Lipid Systems. *J Chem Theory Comput* **2014,** *10* (10), 4730-4744.
61. Abraham, M. J.; Murtola, T.; Schulz, R.; Páll, S.; Smith, J. C.; Hess, B.; Lindahl, E., GROMACS: High performance molecular simulations through multi-level parallelism from laptops to supercomputers. *SoftwareX* **2015,** *1-2*, 19-25.
62. Hoover, W. G., Canonical dynamics: Equilibrium phase-space distributions. *Phys Rev A Gen Phys* **1985,** *31* (3), 1695-1697.
63. Jo, S.; Lim, J. B.; Klauda, J. B.; Im, W., CHARMM-GUI Membrane Builder for mixed bilayers and its application to yeast membranes. *Biophys J* **2009,** *97* (1), 50-8.
64. Wu, E. L.; Cheng, X.; Jo, S.; Rui, H.; Song, K. C.; Davila-Contreras, E. M.; Qi, Y.; Lee, J.; Monje-Galvan, V.; Venable, R. M.; Klauda, J. B.; Im, W., CHARMM-GUI Membrane Builder toward realistic biological membrane simulations. *J Comput Chem* **2014,** *35* (27), 1997-2004.
65. Klauda, J. B.; Venable, R. M.; Freites, J. A.; O'Connor, J. W.; Tobias, D. J.; Mondragon-Ramirez, C.; Vorobyov, I.; MacKerell, A. D., Jr.; Pastor, R. W., Update of the CHARMM all-atom additive force field for lipids: validation on six lipid types. *J Phys Chem B* **2010,** *114* (23), 7830-43.
66. Schneider, T.; Stoll, E., Molecular-dynamics study of a three-dimensional one-component model for distortive phase transitions. *Physical Review B* **1978,** *17* (3), 1302-1322.
67. Berendsen, H. J. C.; Postma, J. P. M.; van Gunsteren, W. F.; DiNola, A.; Haak, J. R., Molecular dynamics with coupling to an external bath. *The Journal of Chemical Physics* **1984,** *81* (8), 3684-3690.
68. Plimpton, S., Fast Parallel Algorithms for Short-Range Molecular Dynamics. *Journal of Computational Physics* **1995,** *117* (1), 1-19.
69. Pan, J.; Tristram-Nagle, S.; Kucerka, N.; Nagle, J. F., Temperature dependence of structure, bending rigidity, and bilayer interactions of dioleoylphosphatidylcholine bilayers. *Biophys J* **2008,** *94* (1), 117-24.
70. Venable, R. M.; Brown, F. L. H.; Pastor, R. W., Mechanical properties of lipid bilayers from molecular dynamics simulation. *Chem Phys Lipids* **2015,** *192*, 60-74.
71. Lafrance, C.-P.; Nabet, A.; Prud'homme, R. E.; Pézolet, M., On the relationship between the order parameter and the shape of orientation distributions. *Canadian Journal of Chemistry* **1995,** *73* (9), 1497-1505.
72. Helfrich, W., Elastic properties of lipid bilayers: theory and possible experiments. *Z Naturforsch C* **1973,** *28* (11), 693-703.
73. Canham, P. B., The minimum energy of bending as a possible explanation of the biconcave shape of the human red blood cell. *J Theor Biol* **1970,** *26* (1), 61-81.
74. Brandt, E. G.; Braun, A. R.; Sachs, J. N.; Nagle, J. F.; Edholm, O., Interpretation of fluctuation spectra in lipid bilayer simulations. *Biophys J* **2011,** *100* (9), 2104-11.
75. Watson, M. C.; Brandt, E. G.; Welch, P. M.; Brown, F. L., Determining biomembrane bending rigidities from simulations of modest size. *Phys Rev Lett* **2012,** *109* (2), 028102.
76. Reynwar, B. J.; Illya, G.; Harmandaris, V. A.; Muller, M. M.; Kremer, K.; Deserno, M., Aggregation and vesiculation of membrane proteins by curvature-mediated interactions. *Nature* **2007,** *447* (7143), 461-4.

77. Simunovic, M.; Voth, G. A., Membrane tension controls the assembly of curvature-generating proteins. *Nat Commun* **2015,** *6*, 7219.
78. Mullinax, J. W.; Noid, W. G., Generalized Yvon-Born-Green theory for molecular systems. *Phys Rev Lett* **2009,** *103* (19), 198104.
79. Frauenfelder, H.; Fenimore, P. W.; Chen, G.; McMahon, B. H., Protein folding is slaved to solvent motions. *Proceedings of the National Academy of Sciences* **2006,** *103* (42), 15469-15472.
80. Dahanayake, J. N.; Mitchell-Koch, K. R., How Does Solvation Layer Mobility Affect Protein Structural Dynamics? *Frontiers in Molecular Biosciences* **2018,** *5*.
81. Pfaendtner, J.; De La Cruz, E. M.; Voth, G. A., Actin filament remodeling by actin depolymerization factor/cofilin. *Proc Natl Acad Sci U S A* **2010,** *107* (16), 7299-304.
82. Blood, P. D.; Swenson, R. D.; Voth, G. A., Factors influencing local membrane curvature induction by N-BAR domains as revealed by molecular dynamics simulations. *Biophys J* **2008,** *95* (4), 1866-76.
83. Zhang, Z.; Lu, L.; Noid, W. G.; Krishna, V.; Pfaendtner, J.; Voth, G. A., A systematic methodology for defining coarse-grained sites in large biomolecules. *Biophys J* **2008,** *95* (11), 5073-83.
84. Katsaras, J.; Tristram-Nagle, S.; Liu, Y.; Headrick, R. L.; Fontes, E.; Mason, P. C.; Nagle, J. F., Clarification of the ripple phase of lecithin bilayers using fully hydrated, aligned samples. *Phys Rev E Stat Phys Plasmas Fluids Relat Interdiscip Topics* **2000,** *61* (5 Pt B), 5668-77.
85. Shapley, L. S., 17. A Value for n-Person Games

Contributions to the Theory of Games (AM-28), Volume II. Kuhn, H. W.; Tucker, A. W., Eds. Princeton University Press: 2016; pp 307-318.
86. Manzhos, S.; Carrington, T., Jr., A random-sampling high dimensional model representation neural network for building potential energy surfaces. *J Chem Phys* **2006,** *125* (8), 084109.
87. Malshe, M.; Narulkar, R.; Raff, L. M.; Hagan, M.; Bukkapatnam, S.; Agrawal, P. M.; Komanduri, R., Development of generalized potential-energy surfaces using many-body expansions, neural networks, and moiety energy approximations. *The Journal of Chemical Physics* **2009,** *130* (18), 184102.
88. Behler, J.; Parrinello, M., Generalized neural-network representation of high-dimensional potential-energy surfaces. *Phys Rev Lett* **2007,** *98* (14), 146401.
89. Bartok, A. P.; Payne, M. C.; Kondor, R.; Csanyi, G., Gaussian approximation potentials: the accuracy of quantum mechanics, without the electrons. *Phys Rev Lett* **2010,** *104* (13), 136403.
90. Wang, H.; Zhang, L.; Han, J.; E, W., DeePMD-kit: A deep learning package for many-body potential energy representation and molecular dynamics. *Computer Physics Communications* **2018,** *228*, 178-184.
91. Schutt, K. T.; Sauceda, H. E.; Kindermans, P. J.; Tkatchenko, A.; Muller, K. R., SchNet - A deep learning architecture for molecules and materials. *J Chem Phys* **2018,** *148* (24), 241722.

**TOC Graphic**

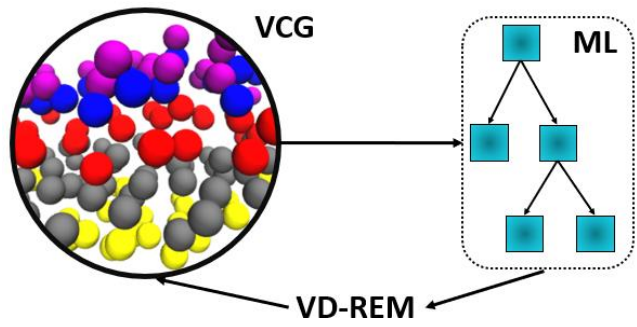

# Supporting Information for

# Utilizing Machine Learning to Greatly Expand the Range and Accuracy of Bottom-Up Coarse-Grained Models Through Virtual Particles

Patrick G. Sahrmann, Timothy D. Loose, Aleksander E.P. Durumeric, and Gregory A. Voth*

*Department of Chemistry, Chicago Center for Theoretical Chemistry, James Franck Institute, and Institute for Biophysical Dynamics, The University of Chicago, Chicago, IL 60637, USA*

*Corresponding Author: gavoth@uchicago,edu

**Description of Machine-Learned Model Features**

As discussed in the main text, features for each machine-learned model consisted of binned statistics of pairwise nonbonded, bonded, and angular interactions, as well as pairwise moments for all interactions. For nonbonded interaction bins, the bin width was set to 0.1 Å for all bins and the outer range for each bin was the pair cutoff of 25 Å. Inner cutoffs were chosen individually through analysis of pairwise statistics. Bonded and angular bins used a bin width of 0.05 Å and 5°, respectively. We list bin ranges for all interactions in Tables S1 and S2.

**Table S1.** Bin ranges for nonbonded and bonded pair interactions.

| Interaction Type | Type 1 | Type 2 | Inner Cutoff (Å) | Outer Cutoff (Å) |
|---|---|---|---|---|
| Nonbonded | HG | HG | 4.0 | 25.0 |
| | HG | MG | 4.0 | 25.0 |
| | HG | T1 | 5.5 | 25.0 |
| | HG | T2 | 7.6 | 25.0 |
| | MG | MG | 4.0 | 25.0 |
| | MG | T1 | 4.0 | 25.0 |
| | MG | T2 | 4.9 | 25.0 |
| | T1 | T1 | 3.4 | 25.0 |
| | T1 | T2 | 3.4 | 25.0 |
| | T2 | T2 | 3.4 | 25.0 |
| Bonded | HG | MG | 3.25 | 7.5 |
| | MG | T1 | 3.5 | 10.0 |
| | T1 | T2 | 3.25 | 10.0 |

**Table S2.** Bin ranges for angular interactions.

| Interaction Type | Type 1 | Type 2 | Type 3 | Inner Cutoff (°) | Outer Cutoff (°) |
|---|---|---|---|---|---|
| Angle | HG | MG | T1 | 40 | 180 |
| | MG | T1 | T2 | 30 | 180 |
| | T1 | MG | T2 | 32 | 179.5 |

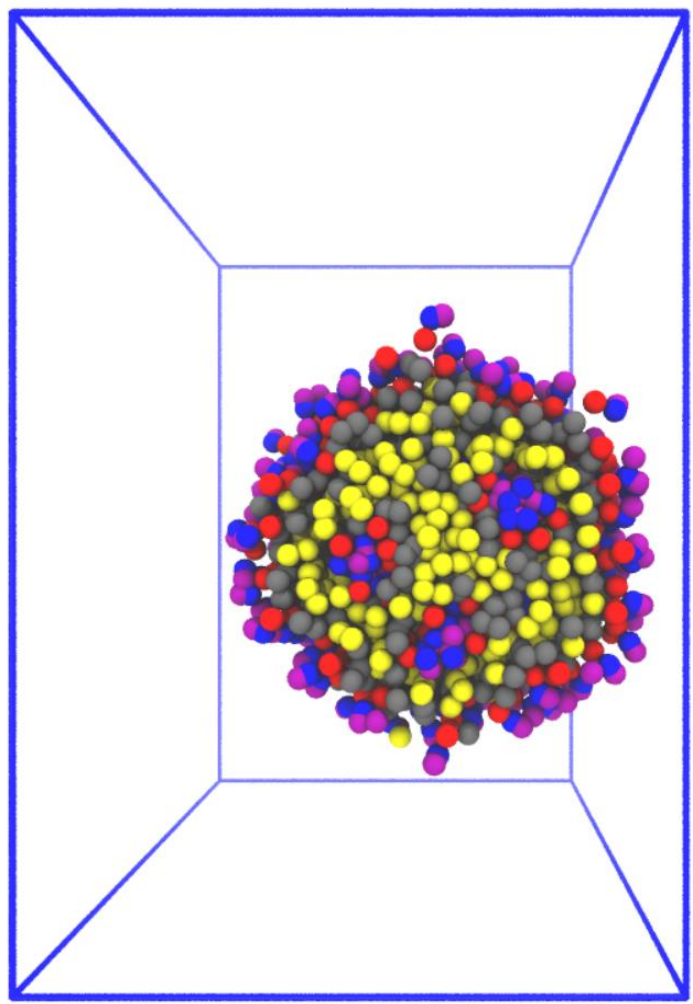

**Figure S1.** Failure of VD-REM-7 to properly self-assemble into a bilayer configuration when virtual particles are rendered non-interacting. Colors are consistent with the main text.

**Movie S1**. Self-assembly of VD-REM-7 from a random, dispersed state. *xy*-dimensions are consistent with reference bilayer area. Simulation occurs over $2 \times 10^6$ MD timesteps. Colors are consistent with the main text.